# GEMINI: A Generic Multi-Modal Natural Interface Framework for Videogames

Luís Filipe Teófilo[1], Pedro Alves Nogueira[1], Pedro Brandão Silva[2]

FEUP – Faculty of Engineering, University of Porto – DEI, Portugal
[1] LIACC – Artificial Intelligence and Computer Science Lab., University of Porto, Portugal
[2] INESC PORTO – Instituto de Engenharia de Sistemas e Computadores, Porto, Portugal
`{pedro.alves.nogueira, lteofilo, pedro.brandao.silva}@fe.up.pt`

**Abstract.** In recent years videogame companies have recognized the role of player engagement as a major factor in user experience and enjoyment. This encouraged a greater investment in new types of game controllers such as the WiiMote™, Rock Band™ instruments and the Kinect™. However, the native software of these controllers was not originally designed to be used in other game applications. This work addresses this issue by building a middleware framework, which maps body poses or voice commands to actions in any game. This not only warrants a more natural and customized user-experience but it also defines an interoperable virtual controller. In this version of the framework, body poses and voice commands are respectively recognized through the Kinect's built-in cameras and microphones. The acquired data is then translated into the native interaction scheme in real time using a lightweight method based on spatial restrictions. The system is also prepared to use Nintendo's Wiimote™ as an auxiliary and unobtrusive gamepad for physically or verbally impractical commands. System validation was performed by analyzing the performance of certain tasks and examining user reports. Both confirmed this approach as a practical and alluring alternative to the game's native interaction scheme. In sum, this framework provides a game-controlling tool that is totally customizable and very flexible, thus expanding the market of game consumers.

**Keywords:** Multi-modal, natural interfaces, videogames.

## 1 Introduction

Videogames and multimedia applications have initially tried to convey increasingly immersive experiences through increased character and environment believability, having in recent years started to dedicate their attention to the interaction artefacts (e.g. the WiiMote™, Kinect™, Move™ and Guitar Hero's controller) [1]. Traditionally, the player is forced to press arbitrary button combinations, which correspond to mapped action in the game world. Often, these combinations are standard (e.g. using the WASD keys to move the game character) or rely on cultural conventions ('R' key for reload, 'F' key for flashlight). Controller-type artefacts, such as the Rock-Band™ instruments, allow players to have a physical mean to interact with the game world.

Natural interaction devices (e.g. Kinect™, voice recognition) allow players to control their avatars by acting as if they were actually performing the task within the game.

Despite natural interaction devices lacking physical means, they allow a much greater range of interaction methods and they also have the ability to integrate them thus proving themselves a more powerful tool for interaction research.

An awarded example of this physical medium/natural interaction fusion is the interactive virtual-reality environment Osmose, by Char Davies [19]. In her experiment, Davies merged a kinaesthetic interaction scheme with traditional virtual reality technologies (a head-mounted display and 3D surround sound) to provide the physical medium. In her own words: "*(Osmose) shuns conventional hand-based modes of user interaction, which tend to reduce the body to that of a disembodied eye and probing hand in favour of an embodying interface which tracks breath and shifting balance, grounding the immersive experience in that participant's own body*" [19]. Davies' results show that some test participants had strong emotional reactions to the whole experience, suggesting that applications reporting high immersion levels (e.g. videogames), coupled with suitable kinaesthetic interaction schemes can drastically increase the enjoyability and sense of emotional engagement of said application. The creation of applications resorting to natural (or kinaesthetic) interaction thus enables more engaging experiences that, in turn may capture the interest of a larger audience and facilitate player engagement along the 4 factors proposed by Lazzaro [2].

Regardless of the increased investment by game designers in crafting more engaging experiences, natural interaction is an area that has not yet been thoroughly researched and thus lacks a set of adequate development tools. This work aims at providing such a tool, usable by both the academic and industry fields. It does so by introducing a versatile framework that can be used to quickly develop and test natural interaction schemes (IS) for applications that were not initially designed for them.

The framework allows the user to define a custom IS via a supplied graphical user interface (GUI) and then use it to interact physically, through recorded poses and verbally, via speech recognition, with the intended application. Poses are automatically recorded and recognised through the user's skeleton, which is detected through a Kinect™ device. Words and sentences are recognized as voice commands by using the Kinect's built-in microphone array and Microsoft's Speech API™ in conjunction. Currently only English is supported. More languages will be added as foreign language libraries become more robust. A dedicated library manages the communication and event handling with the WiiMote™ and Nunckuk™.

## 2 Related Work

### 2.1 Natural Interaction Modes

As previously mentioned, traditional interaction models in videogames resort to button combinations, implemented through keyboard and mouse schemes, only recently shifting to dedicated and natural controllers. However, users are still limited to the designed (native) IS, not being able to redefine or change it altogether. The flexible action and articulated skeleton toolkit (FAAST) is a middleware to simplify the integration of full-body controls with games and virtual reality (VR) applications [3]. To the best of our knowledge it is the only approach that tackles this issue, enabling a toolkit for natural interface implementation. FAAST is able to detect various preset poses and map each one to a single entry of (also preset) actions. Its main limitations are restricting the user to the available preset poses, limiting the action mapping to the preset keyboard and mouse keypress dictionary and not allowing the usage of other interaction devices, other than the Kinect™.

This lack of a standard framework for the deployment of natural interaction schemes leaves individuals researching them with two options. To resort to a Wizard of Oz approach, simulating a non-working prototype [14], which in many cases isn't possible (e.g. playing a game or most real-time activities), or to build his own custom solution from the ground up [15, 16]. The latter is often the only available approach, requiring a huge commitment in terms of time and effort, while also limiting this research field to people versed or with access to people versed in computer science. Additionally, it also stifles the growth rate of the field and its adoption by the public, contributing to its loss of popularity.

### 2.2 Movement Detection

Recent approaches in reliable movement detection have introduced marker-based systems [4], accelerometers [5, 18], physiological sensors [6] and carbon-based strain measurement [17]. While these systems are, in general, accurate they are costly due to the necessary dedicated hardware; and intrusive, by requiring the user to wear the sensors or markers. Some of them also do not measure all of the relevant motions (e.g. strain sensors often do not measure torsion) or provide enough accuracy (e.g. cell phone-grade accelerometers).

With the introduction of the Kinect™ movement detection has become cheap and unobtrusive, alas with some inaccuracy as some of our preliminary tests showed a Gaussian fluctuation of nearly 7 cm on the X and Y planes when the subject was idle. Nevertheless, its cheap price, open source SDK and wide availability induced its use in this study. Despite providing a spatial representation of the user's skeleton, the Kinect™ does not support custom pose or movement recognition. This is an issue that has been vastly studied by the scientific community [7, 8, 9, 10]. Being a complex problem most solutions do not work in real-time or have limited tracking capabilities, which motivated the development of the presented lightweight pose detection method.

### 2.3 Speech Recognition

Speech recognition (SR) is also a complex problem with a multitude of approaches [7, 11, 12, 13]. The main issue with SR is that it requires a database of recognized phonemes and words, which is difficult to create on the fly, as each instance also requires considerable feature extraction and training. Another pressing issue is that it is difficult to identify various sound sources robustly, as well as differentiate from actual sound sources (speakers) and noise. This issue has been tackled by Shih [13], but has yet to be implemented in commercial software. While proving itself resistant to the first issue, the Kinect™ is extremely vulnerable to the two foremost ones at medium distances (~2 meters, the distance required for optimal movement recognition [20]).

Microsoft's Speech API (SAPI) is a widely used package with native support that already provides an extensive database for the English language and features various runtime optimizations, which motivated its choice as our speech recognition engine.

## 3      Conceptual Framework Description

The Generic Multi-Modal Natural Interface (GeMiNI) is a framework meant to support an easy introduction and configuration of any computer compatible peripheral device to work as a game input. It acts as an abstraction layer between device events and the game's default controls. This allows users to experience new interaction methods not originally supported or even been devised by the game's developers. As an example scenario: a first-person shooter game, designed for mouse and keyboard input, could be enhanced using voice commands to trigger actions such as issuing orders to squad members or body poses for crouching, walking or setting traps.

GeMiNI's architecture is conceptually composed of three layers, as depicted in Figure 1. First, the input is captured on the input layer, generating an event type. Different devices with different software drivers (for example, cameras) may output the same type of event (e.g. a captured video frame). The logic layer then translates these into commands that are recognized by the game, according to a user defined IS. More precisely, this is achieved by mapping each event to the game's original controls (e.g. keyboard shortcuts). Lastly, the application layer is responsible to assure that, while in-game, the game actions are invoked when the corresponding events are triggered.

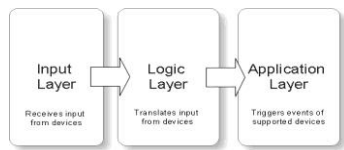

Fig. 1. GeMiNI conceptual Architecture.

## 4      Implementation

The GeMiNI framework is not conceptually restricted to any specific input device. Still, addressing all possible interaction technologies has neither been considered feasible nor relevant at this point of our research. Instead, as a first step, three kinds of device technologies and corresponding events have been considered: Microsoft Kinect cameras, for detecting body poses; embedded microphones for capturing voice commands, and Nintendo Nunchuks for capturing auxiliary key inputs.

For this first approach, these technologies were deemed diverse enough to generate a wide variety of IS. Also, bearing in mind their popularity, maturity and affordability, they were considered the ones where gamers would be more familiarized with and eager to experiment on. The next section addresses more technical details behind the integration of said devices with the proposed architecture's concept and operation.

### 4.1 Architectural Integration

The integration of the three interaction device technologies has been performed according to the conceptual architecture explained in the previous section (see Figure 2). Furthermore, to support an easy addition of new features, each layer is divided into independent modules. As such, an interfacing component was built for each device.

For the implemented devices, three components have been developed in the Input Layer (see Figure 2). The Skeleton Module uses the Microsoft Kinect SDK to process a human skeleton structure from the Kinect camera input, and feeds it to the Pose Recognizer module in the Logic Layer. The Speech Module uses the Microsoft SAPI to capture audio from the microphone, and feeds it to the Speech Recognizer in the Logic Layer. Nunchuk Module resorts to the open-source library WiimoteLib and feeds Nunchuk inputs to the general Input Manager in the Logic Layer.

The Logic Layer then processes the input events using the following components:

— Pose Recognizer: Uses our algorithm based on predefined spatial constraints to detect skeleton poses, and passes an identifier of the pose to the Input Manager;
— Speech Recognizer: Uses the Microsoft SAPI to recognize a designated vocabulary from the audio input, and sends the identified words to the input manager;
— Input Manager: Processes specific identifiers, such as corresponding to keystrokes (from the Nunchuk), to pose (from the pose recognizer) and pronounced words (from the speech recognizer). Then it translates these identifiers to specific game actions, according to a predefined mapping description.

Lastly, the Application layer is connected to the Logic layer and is composed of two distinct modules: The Configuration GUI, which allows an expeditious and intuitive configuration of all the necessary parameters of the architecture's components through a graphical user interface; and the External Application, which is executed simultaneously with the framework and reacts to the translated GeMiNI events.

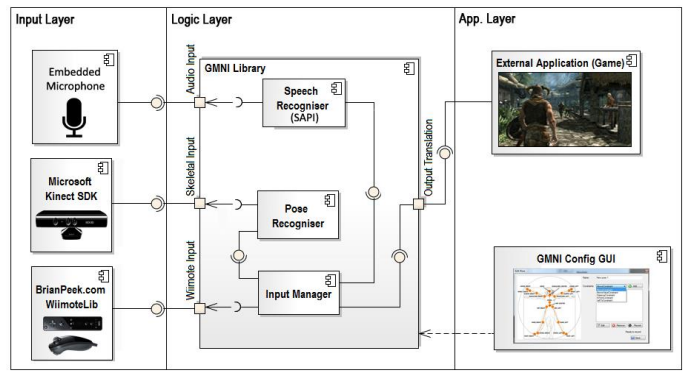

Fig. 2. GeMiNI Implemented Architecture.

### 4.2 Architecture Components

In this sub-section we describe the various components that compose the proposed framework by outlining their features, performance ratings and possible limitations.

**Pose Recognition:** The skeleton data provided by the Kinect cameras contains the position of 20 distinct points (or joints) from the detected human body (see Figure 3).

Each point is imbued with a semantic identifier indicating the body part and a spatial reference in three-dimensional Cartesian space, relative to the cameras. This introduces the possibility to perform queries concerning the relative location of any body part towards another and its 'absolute' location with the Kinect as the origin. The used coordinate system is shown in Figure 4.

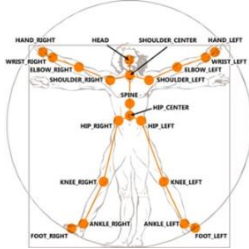 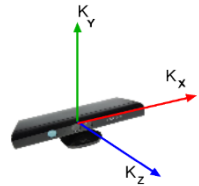

Fig. 3. Skeleton joint disposition[1]     Fig. 4. Kinect Camera coordinate system[2]

Saving a pose by specifying precise absolute location of a joint is neither handy nor feasible, since that it greatly limits possibilities and oversees variations that occur naturally in body posing. Instead, in this approach, poses are specified through sets of spatial constraints, which are verified on body member dispositions. For example, to verify if someone is standing on one foot it is only necessary to consult the Y-coordinate values of the feet joints to check for any prominent differences. Another example: to check if the hands are touching, it is only necessary to calculate the distance between them. This approach allows for a greater flexibility on defining poses.

At this point, the following constraint types have been implemented:

− Distance: Imposes minimum and maximum Euclidean distance between joints;
− InFront: Defines whether a joint is in front of another, by comparing the values of the Z coordinates;
− LeftTo: Defines whether a joint is to the left of another, by comparing the values of the X coordinates;
− AboveOf: Checks if a joint is located above another, by comparing the values of the Y coordinates;
− AboveValue: Checks if the Y coordinate of a joint is located above a certain threshold.

In order to calibrate the pose detection, each constraint can be configured by fine-tuning the corresponding parameters. Still, single constraints may lead to unwanted body pose recognitions since they impose rather broad definitions. In order to reduce this ambiguity it is possible to resort to more than one constraint simultaneously, so as to define a pose. It is important to stress that this constraint-based approach has proven to be fast, so as to allow real-time detection, with an approximate 16-25ms delay, maintaining the game's frame rate.

---
[1] Found at: http://embodied.waag.org/wp-content/uploads/2011/ 07/kinect_joints.png
[2] Adapted from: http://cvcinema.blogs.upv.es/files/2011/05/ convert-kinect-to-standarized1.png

**Pose Recording:** The correct choice and parameterization of the spatial constraints, to define a certain pose, can become a difficult endeavour. For this reason, an alternative to this manual trial-and-error approach has been created. GeMiNI provides completely automatic constraint and parameter definition though pose recording. The Kinect skeleton feed was used to analyse the body's main motion axes during a short (5 second) training phase and infer the relevant constraints and parameters. The method works by the set of joints located in the body's main motion axes (knees, feet, spine, neck, hands and head). Then, it performs peak removal by employing a median filter across a 1.5 second sliding window, with a 0.5 second overlap. This step is done to remove random fluctuations from the recorded signal, as we found the Kinect camera has ~7cm fluctuations on the X and Y planes. The method proceeds to analyse the relations between each possible (distinct) joint pair. For each of these joint pairs, we consider all the possible linear combinations of axes (i.e. X, Y, Z, XY, YZ, XZ and XYZ) and consider that a relevant movement was seen if the maximum observed difference between the joint pair in the recorded data is bigger than a set threshold (15cm in our experiments). Finally, the restriction set is defined as the relevant axis combinations in the analyzed joint pairs.

**WiiMote Communication:** The use of the WiimoteLib to access the Nunchuk inputs has greatly eased the key interpretation, releasing from the need of additional processing in a separate logic layer component. The choice of a Nunchunk driver implementation was motivated by its button diversity (it has both normal buttons and a D-Pad), popularity and compact form. This was considered a good practical alternative for 2D movement or camera control in 3D applications.

**Speech Recognition:** Speech recognition features were implemented, allowing any (pronounceable) word or sentence recognition. The system works at a maximum optimal distance of 2 meters and performs speech recognition with a 1 to 2 second delay. Issues found with SAPI included various user identification and noise cancellation, with some sounds from the environment sometimes being misinterpreted (false positives) as voice commands.

**Input Management and Simulation:** The defined poses, speech commands or external game device outputs can be mapped to a combination of both mouse and keyboard events. Regarding keyboard invocations, there are two event possibilities: Keyboard press, corresponding to a single key-stroke; Keyboard hold, equivalent to holding the key down for certain period, repeating single key-strokes with a certain, configurable frequency. Likewise, mice controls can be simulated in the following manners: Mouse Movement, simulating horizontal and vertical movements; Mouse button press and hold, following the keyboard example.

## 5 Tests & Validation

For the accuracy and usability tests presented in this section, 25 individuals with ages between 18 and 27 years, 76% male and 24% female with no known physical or mental limitations were recruited. Out of these 25 test subjects, 40% were casual gamers - i.e. reported playing less than 1 hour per day and occasionally whenever big titles are released, also having little to moderate familiarity with videogames in general. The remaining 60% were hard-core gamers – i.e. played an average of 3 or more hours per day and possessed advanced videogame and software applications familiarity.

### 5.1 Pose Detection & Inference Accuracy

All twenty-five test subjects were asked to perform twenty designated poses, which were recorded and automatically inferred by GeMiNI. Subjects were verbally instructed so as not to condition their interpretation of the required poses, thus enabling the most natural experience possible and also testing the method's robustness. They were then asked to re-enact each one of these poses ten times, to measure the detection accuracy. The poses' inference accuracy was also considered in these tests. The inference of poses is the process of automatically determining the constraints that characterize a given pose. We considered a pose learned if, after the inference process, it is detected with an accuracy of 80%, or more, on subsequent repetitions. This value was empirically defined as we found subjects reported frustration with the system bellow this accuracy threshold. Each pose was repeated 20 times by each participant. Overall results for the test population are depicted in Table 1.

Table 1. Average pose accuracy detection and inference (D – Detection, I – Inferring)

| Pose | D | I | Pose | D | I |
|---|---|---|---|---|---|
| Step forwards/backwards | 97% | 95% | Crouch | 76% | 92% |
| Lean left/right | 81% | 90% | Flex | 93% | 100% |
| Left/right punch | 99% | 100% | Point a bow | 94% | 94% |
| Lift left/right leg (kick) | 97% | 100% | Hands behind shoulders | 86% | 87% |
| Jump | 96% | 96% | Outstretched arms | 95% | 96% |
| Raise left/right arm | 100% | 100% | Lean forwards/backwards | 86% | 87% |
| Grabbing motion | 98% | 96% | | | |

### 5.2 Speech Recognition

For this test, fifty words were randomly selected from the game's item inventory list. These words were also used to generate fifty sentences composed of two or three of those words (as a voice command is usually comprised of one to three words). The set of sentences was then segmented into three complexity categories according to the number of syllabi of the sentences. Thus, these categories represent the simplest to most complex voice commands possible in our test scenario.

Each test subject then repeated a set of 15 random samples from each category 5 times, in order to test the SR accuracy. The tests were performed at a distance of about 2 meters - the optimal distance for body movement capture [20] - from the microphone in a quiet room. The accuracy for this task is depicted in Table 2.

Table 2. Speech recognition accuracy results.

| Syllabae count | Single words | Sentences |
|---|---|---|
| 1-3 syllabae | 78% | 89% |
| 3-5 syllabae | 83% | 93% |
| 6-8 syllabae | N/A | 94% |
| 8-10 syllabae | N/A | 93% |

### 5.3 Usability Testing

To evaluate GeMiNI's GUI usability, each of the test subjects were asked to perform a series of tasks (Table 3) that represented each of the previously mentioned steps involved in defining a new IS. The mean values and standard deviations for each action's completion times were calculated, as well as the total number of errors performed by the subjects. These results are present in Table 3.

Table 3. Task completion times and errors.

| Task | Mean (sec) | Stand. Dev. | Total Errors |
|---|---|---|---|
| New pose (manual) | 46.4 | 14.7 | 254 |
| New pose (auto) | 17.4 | 5.2 | 23 |
| New voice command | 8.6 | 3.5 | 9 |
| Add simple action | 13.7 | 4.8 | 29 |
| Add complex action | 32.6 | 13.3 | 99 |
| Set Wiimote button | 4.1 | 2.3 | 13 |

## 6 Case Study – Bethesda's Elder Scrolls V: Skyrim

Despite having developed ISs for various games (e.g. Devil May Cry 4, Super Mario, Legend of Zelda) for our case study we wanted to focus on a videogame with complex and diverse interaction mechanics. Thus, we sought a videogame that provided seamless and complex combat, social and user interface interaction, ultimately choosing the videogame: The Elder Scrolls V: Skyrim. Skyrim is an open-world action role playing game where the player must explore dangerous locations and interact with various factions to further the storyline. Given its genre, the game allows virtually unlimited freedom of movement and interaction through dozens of gameplay mechanics, thus proving an alluring test bed for our framework.

### 6.1 Interaction Scheme

The game's IS is mainly divided into three categories: exploration, social interaction and combat (exemplified in Figures 6 and 7). Table 4 shows the selected test actions and the relation between their native and newly defined IS.

Table 4. Comparison between native (Nat-IS) and new IS (New-IS).

| Action | Nat-IS | New-IS |
| --- | --- | --- |
| Move forward/backward | 'W/S' | Right foot forward/backward more than 20cm |
| Strafe left/right | 'A/D' | Lean left/right more than 20º |
| Orientation (look around) | Mouse cursor | Move Wii Nunchuk in desired direction |
| Invoke map | 'M' | Outstretched arms |
| Initiate a conversation | 'E'' | Wave/Say 'hello' |
| Quit a conversation | 'Tab' | Say 'goodbye' or 'see you soon' |
| Buy/sell an item | 'Enter' | Say 'buy/sell' |
| Equip weapon/spell | '1-8' | Say weapon/spell name |
| Use weapon / cast spell | Mouse click | Push equipped hand forward |
| Charge spell | Mouse hold | Raise corresponding arm |
| Raise shield | Right mouse click | Arm in front of chest with horizontal orientation |
| Charge at enemy | 'Alt + W' | Right foot forwards more than 30 cm |

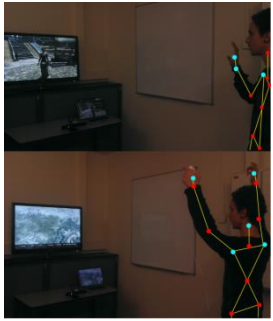

Fig. 5. The "cast spell" and "raise shield" actions, respectively.

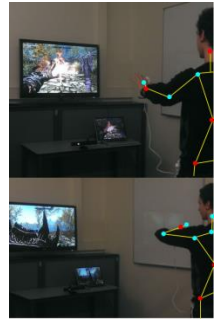

Fig. 6. The "initiate a conversation" and "invoke map" actions, respectively.

### 6.2 User Commentaries

A pilot study with sixteen test subjects with no previous knowledge of the video game's commands was carried out. Each subject played the game with the native and the natural interaction scheme. Afterwards, players were asked to answer a brief questionnaire, so as to gather data about their preferences. The questionnaire contemplated the three interaction categories. Each of the test subjects was asked to evaluate which scheme they preferred and rate each IS. Results (Table 5) show that users found the natural scheme to provide a more enjoyable and intuitive user-experience.

Besides the aforementioned evaluation method, the subjects were also asked to assess the natural scheme qualitatively. The subjects highlighted that the system's response time (16-25 *ms*) was adequate. They also gave special emphasis to the possibility of customizing the interaction scheme. Finally, test subjects pointed out that due to the smaller amount of mix-ups in the natural interface, the learning curve becomes faster. This suggests that the games' first impression on users with a natural interface improves considerably thus creating a more addictive experience. Further studies are required to correctly assess the truthfulness of this statement and quantify how much faster the learning curve actually is.

Table 5. Interaction scheme user preferences. Overall, users preferred the natural one.

| Game feature | Prefer Native | Prefer Natural | No Preference |
|---|---|---|---|
| Social Interaction | 25% | 50% | 25% |
| Inventory Management | 19% | 37% | 44% |
| Movement / Exploration | 19% | 56% | 25% |
| Combat | 13% | 81% | 6% |

## 7  Conclusions & Future Work

In comparison with existing works, this approach is faster and provides more features, such as new pose definitions and their automatic recording, support for other devices, speech recognition and complex input mappings without the need for third-party software. However two technical limitations were found. Firstly, that the Kinect™ must be distant from the speakers so as to not interpret voices or sounds coming from the game as voice inputs. Secondly, some actions (e.g. shaking someone's hand, opening a door and casting a spell) require some form of context to be correctly identified, which is undoable without access to the game's engine. In other words, it only shows any limitations when the Kinect™ sensor was poorly placed or when used with complex applications that did not previously support natural interaction.

Retrospectively, the system has proved itself capable of delivering an accurate, versatile and satisfactory method for the implementation of multi-modal natural interfaces, as has been proved by our trials. It also succeeded in providing a pleasurable experience in one of the most complex action videogames currently available.

Future work should focus on performing a set of comprehensive immersion studies on a larger population so as to quantify how much physical involvement can benefit the overall experience. Head tracking could be used as an interaction mode or to gather involvement data. Finally, a streamlined version of the GUI could also be created so as to make the framework available to an even broader population.